\documentclass[%
 reprint,
superscriptaddress,
 amsmath,amssymb,
 aps,
]{revtex4-1}

\usepackage{graphicx}
\usepackage{dcolumn}
\usepackage{bm}
\usepackage{color} 
\usepackage{ulem}

\begin{document}
\preprint{APS/123-QED} 
\title{Direct observation of spatio-temporal dynamics of short electron bunches in storage rings }


\newcommand{\affiliationSOLEIL}{\affiliation{Synchrotron SOLEIL, Saint Aubin, BP 34, 91192 Gif-sur-Yvette, France}}
\newcommand{\affiliationPhLAM}{\affiliation{Laboratoire de Physique des Lasers, Atomes et Mol\'ecules, UMR CNRS 8523\\ Centre d'\'Etudes et de Recherches Lasers et Applications, FR CNRS 2416,   Universit\'e des Sciences et Technologies de Lille, F-59655 Villeneuve d'Ascq Cedex, France}}
\newcommand{\affiliationElettra}{\affiliation{Elettra-Sincrotrone Trieste, Strada Statale 14-km 163,5 in AREA Science Park,34149 Basovizza, Trieste, Italy}}

\author{C. Evain}\email{clement.evain@univ-lille1.fr}  \affiliationPhLAM 
\author{E. Roussel}\affiliationElettra
\author{M. Le Parquier}\affiliationPhLAM 
\author{C. Szwaj}\affiliationPhLAM 
\author{M.-A. Tordeux}\affiliationSOLEIL
\author{L. Manceron}\affiliationSOLEIL
\author{J.-B. Brubach}\affiliationSOLEIL
\author{P. Roy}\affiliationSOLEIL
\author{S. Bielawski}\affiliationPhLAM

\date{\today}

\begin{abstract}
In recent synchrotron radiation facilities, the use of short (picosecond) electron bunches is a powerful method for producing giant pulses of Terahertz Coherent Synchrotron Radiation (THz CSR). Here we
report on the first direct observation of these pulse shapes with a few picoseconds resolution, and of their dynamics over a long time. We thus confirm in a very direct way the theories predicting an interplay between two physical processes. Below a critical bunch charge, we observe a train of identical THz pulses (a broadband Terahertz comb) stemming from the shortness of the electron bunches. Above this threshold, a large part of the emission is dominated by drifting structures, which appear through spontaneous self-organization. These challenging single-shot THz recordings are made possible by using a recently developed photonic time stretch detector with a high sensitivity. The experiment has been realized at the SOLEIL storage ring. 
\begin{description}
\item[PACS numbers]41.60.Ap, 29.27.Bd, 42.65.Re, 07.57.Hm
\end{description}
\end{abstract}

\pacs{41.60.Ap, 29.27.Bd, 42.65.Re, 07.57.Hm}

\maketitle

In last generation storage ring facilities, it has become possible to produce pulses of Coherent Synchrotron Radiation (CSR), with powers that exceed “classical” (incoherent) synchrotron radiation by factors greater than 10000~\cite{Carr2001, Abo-Bakr2002, Byrd2002,  Mochihashi2006, Karantzoulis2010, Barros2013, Billinghurst2015}, and may be organized in Terahertz combs~\cite{Steinmann2015,Tammaro2015}.

\begin{figure*}
 \includegraphics[width=0.88\textwidth]{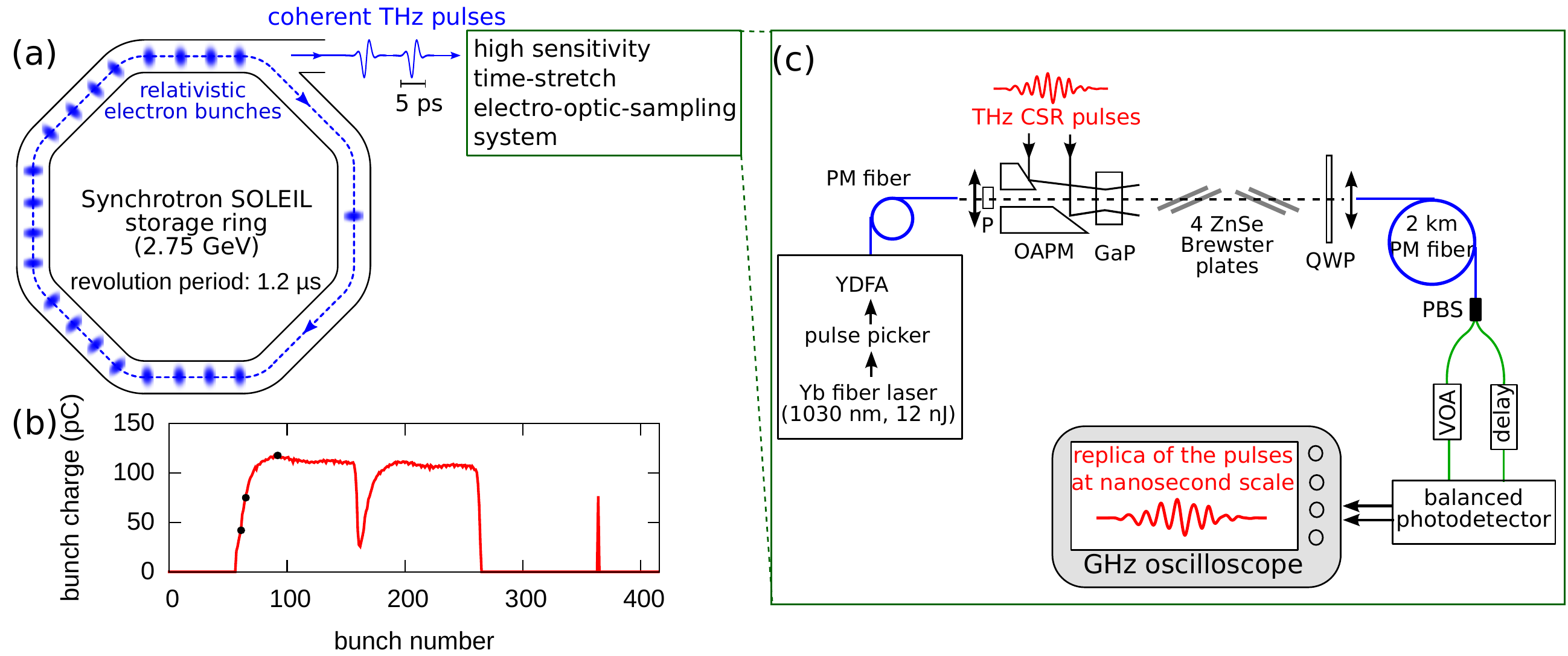}
  \caption{a) Global experimental arrangement.  The SOLEIL storage ring is operated with many short electron bunches (209 in the experiment). The THz coherent synchrotron radiation pulses (CSR) emitted in a bending magnet are recorded in single-shot with picosecond resolution at the THz/IR AILES beamline, using high-sensitivity photonic time-strech electro-optics sampling.  (b) Distribution of the electron bunches charges over one turn for this experiment. The three dots indicate the electron bunches whose CSR pulses evolutions are displayed in Figures~\ref{fig:shots} and \ref{fig:exp_2d}.  (c) photonic time stetch electro-optical sampling setup. Blue lines: polarization-maintaining (PM) fibers (PM980). green lines lines: single mode fibers (HI1060). PBS: fibered polarizing beam-splitter. OAPM: Off-Axis parabolic mirror, P: polarizer, GaP: 5 mm long, [110]-cut Gallium Phosphide crystal, QWP: quarter-wave plate, VOA: variable optical attenuator, “delay”: adjustable delay line. The balanced detector is an DCS-R412 from Discovery Semiconductors. The use of the four ZnSe plates enables to enhance the setup detectivity (see text and Ref.~\cite{Ahmed2014}). The polarizations of the laser and THz field are parallel and oriented along the [-110] axis of the GaP crystal  (and peremptory to the Figure plane). The QWP is adjusted at 45  degrees with respect to the laser polarization.}
    \label{fig:setup}
\end{figure*}

A priviledged way for obtaining this type of emission with the stability required for users, consists in operating the storage ring facilities with short (picosecond) electron bunches~\cite{Abo-Bakr2002,  Abo-Bakr2003, Mathis2003,  Feikes2011, Martin2011,  Barros2013, Billinghurst2015, Barros2015}. 
As the main idea, if the longitudinal shape of an electron bunch presents Fourier components at terahertz frequencies, an intense and stable emission of coherent terahertz radiation is generally expected in bending magnets of accelerators, as illustrated in Figure~\ref{fig:setup}.

This simple conceptual picture is nevertheless deeply complicated by a physical phenomenon of very fundamental nature: the interaction of electron bunches with their own emitted radiation leads to a spatiotemporal instability, which is characterized by the spontaneous appearance of microstructures at the millimeter scale~\cite{Stupakov2002, Venturini2002, Roussel2014a}. In turn, these structures also emit CSR. Hence the interplay between these two modes of emission (i.e., from the "shortness" of the electron bunches and from the spontaneously formed microstructures) plays a central role in the knowledge and improvement of the present and future CSR sources.

However, it is only recently that direct experimental observation of the emitted THz electric field pulse shapes became possible.
As an important milestone, electro-optic sampling has become possible in storage rings either for recording CSR pulses~\cite{Katayama2012, Mueller2012}, and electron bunch near-field shapes~\cite{Hiller2014}. Then the so-called photonic time-stretch strategy \cite{Coppinger1999, Goda2013} opened the way to direct recordings of CSR electric field pulse shapes, over successive shots at ultra-high repetition rates~\cite{Roussel2015}.

Nevertheless, the limited sensitivity of these new high-repetition-rate picosecond detectors limited experimental investigation of THz signals to special conditions of high bunch charge and long electron bunches. This configuration corresponds to highest possible peak powers, but also to strongly unstable operation~\cite{Roussel2014a, Roussel2015}. Thus, characterization of CSR pulse shape dynamics remained an open question for short (picosecond-scale) and low-charge electron bunches, i.e.,  the specific conditions where exploiting this fundamental phenomenon, as a THz source, are foreseen. 

In this Letter, we present the actual shapes (the electric field evolution, including envelope and carrier) of the Terahertz CSR pulses emitted in short bunch operation in a storage ring. This analysis has been possible by combining two recent advances in Terahertz electric field detection: high-sensitivity electro-optic sampling~\cite{Ahmed2014, Savolainen2013} for obtaining the required signal-to-noise ratio with picosecond resolution, and the so-called time-stretch strategy~\cite{Coppinger1999, Goda2013, Roussel2015}, which enables single-shot recordings at ultra-high acquisition rates. As a main result, we present a direct characterization of the transition from stable Terahertz emission to dynamically evolving pulses: the microbunching instablity threshold, which plays a central role in theories of CSR~\cite{Venturini2002, Stupakov2002}. We also show that numerical integration of the Vlasov-Fokker-Planck equation reproduces the main characteristics of the experimental signals.\\

The experiment has been performed at the SOLEIL storage ring in a so-called {\it low-alpha} user shift. This mode is characterized by the operation of short bunches (of about 5~ps RMS~\cite{Tordeux2012}), which emit CSR with a stability which is compatible with THz user experiments. Moreover the 209 bunches which circulate in the storage ring have typically different charges [Fig.~\ref{fig:setup}(b)].
This allows investigations as a function of bunch charge by simply studying the emission of each individual electron bunches, without change of the storage ring operating conditions.

\begin{figure}[htbp!]
  \includegraphics[width=1\linewidth]{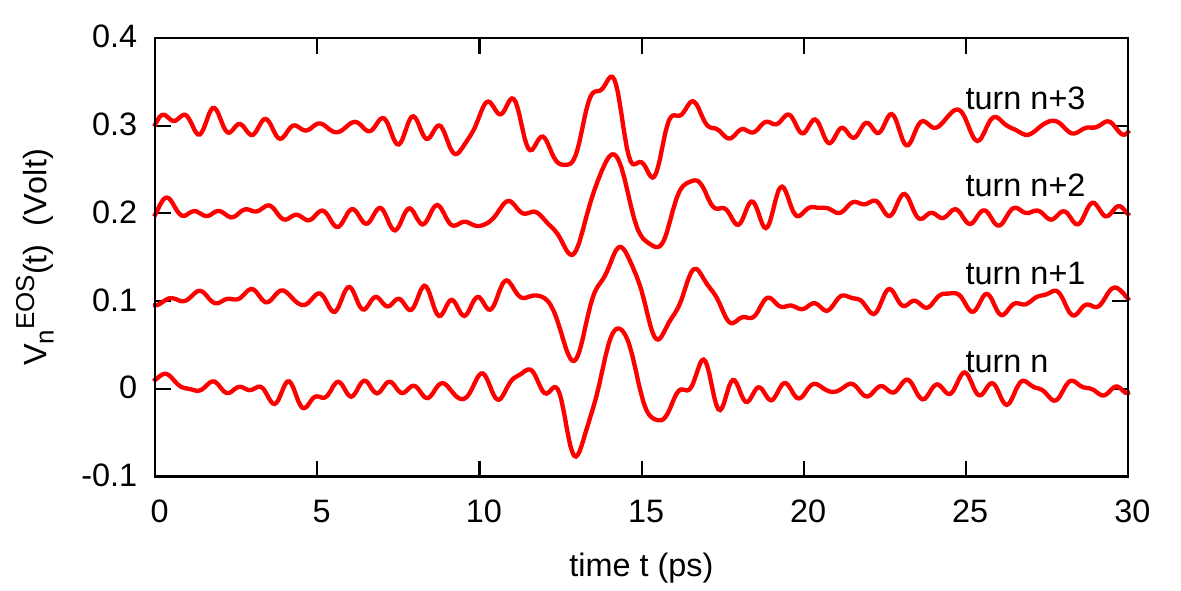} 
  \caption{Typical single-shot electro-optic sampling traces $V_n^{EOS}(t)$ (corresponding to the THz CSR electric field emitted by one electron bunch, at four successive turns in the storage ring).  The electron charge is 75~pC [second dot, i.e bunch number \#65 on Figure~\ref{fig:setup}(b)]. For clarity,  the signals are shifted vertically by 0.1~V.}
  \label{fig:shots}
\end{figure}

\begin{figure*}
  \includegraphics[width=\textwidth]{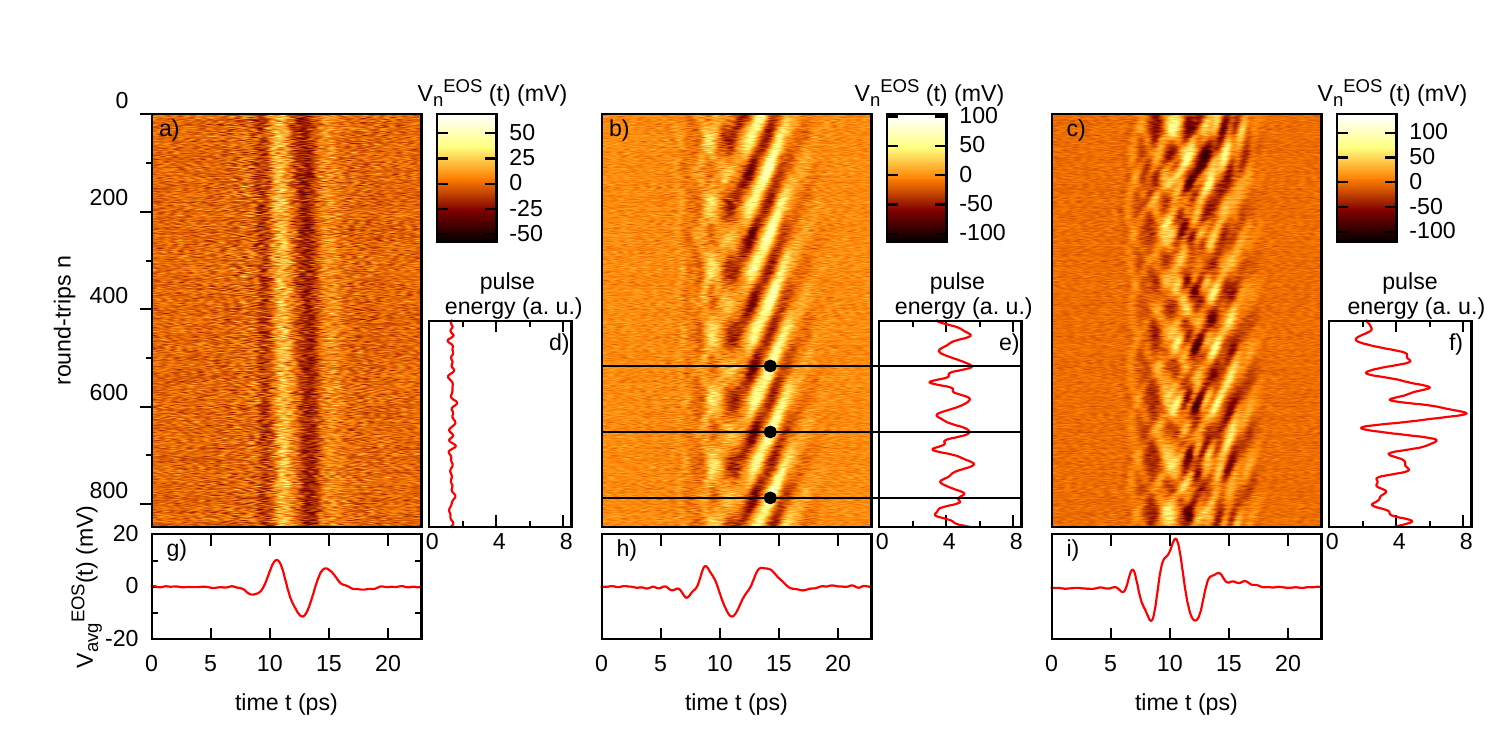}
  \caption{(a)-(c): Terahertz pulses shapes (electro-optic sampling signal $V_n^{EOS}(t)$) versus round-trips $n$ in the storage ring. below and above the microbunching instability threshold. (a)-(c) correspond to bunch charges of Q=42~pC, 75~pC and 118~pC respectively [i.e, bunch numbers \#61, \#65, and \#92  in Fig.~\ref{fig:setup}(b)]. Each horizontal cut corresponds   to the electric field shape (as in Fig.~\ref{fig:shots}).   (d)-(f): pulse energy versus turn number (integration over the  time of the square of the EOS signal, and low-pass filtered at 50 kHz).  Horizontal black lines and  dots placed for visual guide. (g)-(i): average electro-optic signal $V_{avg}^{EOS}(t)$.  Note that the data  (a) and (b)  are recorded simultaneously (i.e., they correspond to a       single oscilloscope recording).}
  \label{fig:exp_2d}
\end{figure*}
The THz CSR pulses are recorded on the AILES beamline, using a photonic time-stretch electro-optic sampling setup, which is specially designed for high sensitivity [Fig.~\ref{fig:setup}(c)]. This detection system is composed of two parts, which are partly intertwined (see also Ref.~\cite{Szwaj2016} for details). The first part is a single-shot EOS~\cite{Jiang1998,Wilke2002} system based on the design developped at SLS~\cite{Mueller2012}, FLASH~\cite{Steffen2009} and ANKA~\cite{Hiller2014},  which imprints the THz electric evolution onto the longitudinal profile of a chirped laser pulse at 1030~nm. As a key point for  reaching the required high sensitivity, we follow the special  strategy recently published by Ahmed, Savolainen and Hamm~\cite{Ahmed2014, Savolainen2013}, which consists in using   balanced detection, and introducting a set of Brewster plates as  shown in Figure~\ref{fig:setup}(c). In order to obtain the necessary high  acquisition rates, we associated this EOS setup with the so-called photonic time stretch technique developed by B.~Jalali and coworkers~\cite{Coppinger1999,Wong2011}. The chirped laser pulses containing the THz pulse shape information are stretched within a 2 km-long fiber~\cite{Goda2013}, up to the nanosecond scale. At the balanced photodetector’s output, we thus obtain a "replica" of CSR  pulse, that is temporally magnified, by a factor of about 200.
The optical performances of the setup are described in detail in Ref.~\cite{Szwaj2016}. The acquisition window is of the order of 15~ps, and the sensitivity is of 2.5~$\mu$V/cm/$\sqrt{\textnormal{Hz}}$ at 300~GHz in the electro-optic crystal (i.e., 1.25~V/cm). Since the frequency range of the CSR pulses is typically in the 150-600 GHz range~\cite{Barros2013, Barros2015}, the “stretched signal” to be recorded by the oscilloscope is in the 0.75-3~GHz range. Here we use a Teledyne LeCroy LabMaster 10-65Zi oscilloscope (with an -- overdimensioned -- 30 GHz bandwidth), and the acquired data are numerically low-pass filtered at 6 GHz. We recorded 3 ms long time series, which corresponds to approximately 2500 turns in the storage ring.
Typical single-shot recordings of the THz CSR pulses (electric field versus time) are represented in  Figure~\ref{fig:shots}. In order to obtain a synthetic view of the  CSR pulse dynamics over a large number of turns in the storage ring, we also  represented the pulse series in a colorscale diagram, versus time  and number of turns, as displayed in Figure~\ref{fig:exp_2d}(a)-(c),  for three values of bunch charge.

 These diagrams clearly reveal two different emission processes.
First, the data evidence the coherent emission which was predicted to occur from the ``shortness'' of the electron bunch~\cite{Sannibale2004}. This process is directly visible below the microbunching instability threshold [Fig.~\ref{fig:exp_2d}(a)]. The emission occurs as a few-period coherent THz pulses, without dynamical evolution from turn-to-turn. This corresponds to a quasi-perfect Terahertz frequency comb~\cite{Tammaro2015}.  

A second process occurs when the charge exceeds the so-called microbunching instability threshold [Figs.~\ref{fig:exp_2d}(b,c)].  An important part of the THz radiation is then emitted in the form of an oscillation, that drifts at each turn in the storage ring. This carrier-envelope phase drift stems from the appearance of microstructures which are continuously moving inside the electron bunch, as described in the numerical simulations presented later.
In addition, the first process, i.e. the coherent emission due to the bunch shortness, is expected to exist above the microbunching instability threshold. 
The signature of this contribution from the data presented in Figs.~\ref{fig:exp_2d}(b,c) can be evidenced by computing the electric field pulse average over a large number of turns: 
\begin{equation}
  V_{avg}^{EOS}(t)=\sum_{n=n0}^{n0+N-1}V_{n}^{EOS}(t),
\end{equation}
where $V_{n}^{EOS}(t)$ is the electro-optic sampling signal of the  THz pulses. This processing filters out the drifting component and may thus be viewed as a measure of the component due to the bunch shortness. This  contribution $V_{avg}^{EOS}(t)$ (with $N\simeq 850$) is visible in  Figures~\ref{fig:exp_2d}(g,h,i). Note that this component is linked  to the short-bunch mode operation, i.e., is not observed when long  electron bunches are used~\cite{Roussel2014a,Roussel2015}. 

From these data, it is also possible to reconstruct the turn-by-turn evolution of the THz pulse energy  $E_{THz}^n$, with 
\begin{equation}
 E_{THz}^n=\int_{t_1}^{t_2} |V_{n}^{EOS}(t)|^2 dt,
\end{equation}
where $t_1$ and $t_2$ are taken near the boundaries of the THz pulse. 
This signal corresponds to the signal that would be recorded with a standart "slow" detector [see Figs.~\ref{fig:exp_2d}(d-f)].
It permits to confirm in a very direct way  that the modulation in the THz signal, which can be clearly seen in Fig.~\ref{fig:exp_2d}(e) and which had been observed in numerous studies~\cite{Kuske2009, Feikes2011, Evain2012, Roussel2014, Judin2012, Bartolini2011,Billinghurst2016}, is directly related to the period of apparition of the micro-structures in the bunch [Fig.~\ref{fig:exp_2d}(b)]. 

We have then performed numerical integrations of analytic models, to make a comparison with these experimental data (obtained for the first time in a short-bunch mode), and also to access the 2D longitudinal phase-space.
We present here the results obtained with one of the most elementary models: the one-dimensional Vlasov-Fokker-Planck model with shielded CSR wakefield~\cite{Venturini2002}.
The evolution of the electron density is described by the phase space distribution $f(q, p, \theta)$, where $\theta$ is the time associated to the number of round-trips in the ring, and $q$ and $p$ are longitudinal coordinate, and electron energy variables respectively. The dynamical evolution of $f (q, p,\theta)$ is determined by the following Vlasov-Fokker-Planck equation~\cite{Venturini2002}:

\begin{eqnarray} \textstyle
  \frac{\partial f(q,p,\theta)}{\partial \theta} &-& \textstyle  p \frac{\partial f}{\partial q} +  \frac{\partial f}{\partial p} \left( q - I_c E_{wf} \right) = 2\epsilon \frac{\partial}{\partial p}\left( p f + \frac{\partial f }{\partial p}\right)  \label{eq:VFP1}
\end{eqnarray}

The time $\theta$ associated to round-trips is a dimensionless variable: $\theta = 2\pi f_s t$, where $t$ is the time (in seconds) and $f_s$ is the synchrotron frequency. The longitudinal position $q$ and relative momentum $p$ are the deviation from the so-called synchronous electron (with position $z_0$ and energy $E_0$). $q$ and $p$ are expressed in units of the equilibrium bunch length $\sigma_z$ and energy spread $\sigma_E$ at zero current: $q=(z-z_0)/\sigma_z$, and $p = (E - E_0)/\sigma_E$.  $\epsilon = 1/(2π f_s \tau_s )$, where $\tau_s$ is the synchrotron damping time.
$I_c=I \frac{e2\pi R_c}{2\pi f_s \sigma_E T_0}$, with $I$ is the average beam current ($I=\frac{Q}{T_0}$), $R_c$ the dipole radius of curvature, and $T_0$ the revolution period. All parameters are in MKS units.
\begin{figure}[htbp!]
  \includegraphics[width=1\linewidth]{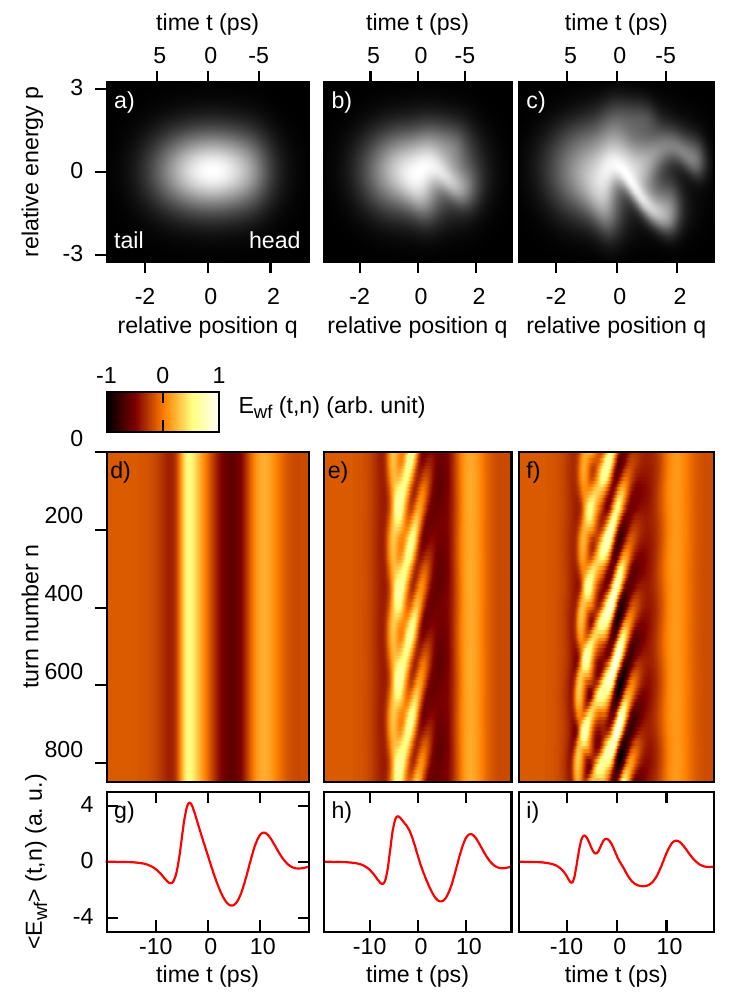}
  \caption{Numerical simulations of the bunch dynamics, with a charge Q of 82.6~pC, 94.4~pC, and 141.6~pC, left, middle and right column  respectively.  (a)-(c): electron distribution in  the phase-space (at the turn number $n\simeq 846$). (d)-(f): evolution of the electric field pulses (inside the chamber) $E_{wf}(t,n)$ versus turn number, with $t=-q\sigma_z/c$, $n=\theta/(2\pi f_s T_0)$, $c$ the light velocity.  (g)-(i): average over the  turns of the emitted pulses.  Simulation parameters:   $E_0=2.75~$GeV, $f_s=928$~Hz, $\sigma_E=E_0 \times 1.017\times 10^{-3}$, $\sigma_z=0.92~$mm, $\tau_s=3.27~$ms, $R_c=5.36$~m, $T_0=1.181~\mu$s,  $h=12.5~$mm.}
  \label{fig:num}
\end{figure}

The interaction between electrons is described by the term $E_{wf}(q,\theta)$ which represents the electron bunch wakefield (expressed in V/m, per Ampere of average current in the storage ring). We use here the CSR wakefield created by an electron bunch in a circular trajectory, between two parallel plates with distance 2h. The exact expression of $E_{wf}(q,\theta)$ is well known~\cite{Murphy1997}, and detailed in~\cite{Roussel2014}. Integration of Eq. (1) is performed using the semi-Lagrangian scheme, described in Ref.~\cite{Warnock2000}, and parameters are listed in the figure~\ref{fig:num} caption. 

Figure~\ref{fig:num} shows the results of the numerical integration of equation~(1) for three bunch charges $Q=82.6~$pC, 94.4~pC and 141.6~pC.
We retreive here qualitatively  important features observed experimentally.
Below the instability threshold, no micro-structures are observable on the phase-space [Fig.~\ref{fig:num}(a)] and the shape of the electron bunch is constant from turn-to-turn [Fig.~\ref{fig:num}(d)]. Just above the instability threshold, "branches" continuously appears on the bottom part of the phase-space, and display a rotational motion at the synchrotron frequency [see Fig.~\ref{fig:num}(b) and supplemental movie]. As a main consequence, the wakefield presents an oscillation which is  drifting toward the bunch head [Fig.~\ref{fig:num}(e)]. For higher  currents, the micro-structures are more prominent [see Fig.~\ref{fig:num}(c) and supplemental movie] and can reach the upper part of phase space. In this case, the wakefield  [Fig.~\ref{fig:num}(f)] also presents an oscillation which is drifting   toward the tail of the bunch (in addition to the oscillation   drifting to the head).

We believe that the availability of this new type of data (CSR pulse shapes versus round-trips) is a precious tool for testing  models of bunch dynamics (including wakefields), in a more clear-cut  way than with classical means (i.e., average spectra, THz pulse energies, microbunching instability threshold, etc.). In the example presented here (where  one of the most elementary model of wakefields is used), agreement   as well as differences can clearly be seen. For instance   below threshold, the wakefield displays a longer period in the   numerical simulation [Fig.~\ref{fig:num}(g)] than experimentally  [Fig.~\ref{fig:exp_2d}(g)]. This may be attributed to differences in the  wakefield modeling and/or in filter effects in the beamline wich are  not taken into account in the modeling. Differences at higher  current are also perceptible, as experimental data  [Fig.~\ref{fig:exp_2d}(c)] are more irregular than numerical ones [Fig.~\ref{fig:num}(f)]. We think that such comparisons of numerical and   experimental diagrams can now be used as an important guide for the   modeling process, including the  crucial  work of  accelerator impedance   modeling.\\
In conclusion, it is now possible to analyze in a complete way   the coherent synchrotron radiation emission due to short electron   bunches in storage rings. As a main result, we could directly visualize   the predicted transition between a stable THz coherent emission (due   to the ``bunch shortness'') and the emission due to the so-called   microbunching instability. In the future, we believe that this viewing of   CSR dynamics enables new and stringent tests for modeling the present and   future storage ring facilities. 
In another perspective, there results show that THz spectra can be obtained at a MHz repetition rate, paving the way for ultra high repetition rate time resolved THz spectroscopy.

We would like to thanks Menlo Systems for important advices. The work has been supported by the BQR of Lille University (2015), by the Ministry of Higher Education and Research, Nord-Pas de Calais Regional Council and European Regional Development Fund (ERDF) through the Contrat de Projets \'Etat-R\'egion (CPER) 20072013, and by the LABEX CEMPI project (ANR-11-LABX-0007).
%

\end{document}